\documentstyle[12pt]{article}
\newcommand{\bra}{\langle}
\newcommand{\ket}{\rangle}
\newcommand{\R}{\mbox{\boldmath $ R $}}
\newcommand{\Z}{\mbox{\boldmath $ Z $}}
\newcommand{\map}{\mbox{Map}}
\newcommand{\diff}{\mbox{Diff}^+}

\makeatletter
\@addtoreset{equation}{section}
\makeatother

\begin{document}
%
%
\begin{flushright}
DPNU-94-60
\\
hep-th/9412174
\\
December 1994
\end{flushright}
\vspace*{12mm}
\baselineskip 10mm
\begin{center}
{\Large \bf %
Zero-mode, winding number and quantization \\
of abelian sigma model in (1+1) dimensions}
\vspace{12mm}
\\
{\large %
Shogo {\sc Tanimura}\footnote{%
e-mail address : tanimura@eken.phys.nagoya-u.ac.jp}
}
\\
{\it Department of Physics, Nagoya University,
Nagoya 464-01, Japan}
\vspace{16mm}
\\
{\bf Abstract}
\\
\begin{minipage}[t]{120mm}
\baselineskip 6mm
We consider the $ U(1) $ sigma model in the two dimensional space-time
$ S^1 \times \R $,
which is a field-theoretical model possessing a nontrivial topology.
It is pointed out that its topological structure is characterized
by the zero-mode and the winding number.
A new type of commutation relations is proposed to quantize the model
respecting the topological nature.
Hilbert spaces are constructed
to be representation spaces of quantum operators.
It is shown that there are an infinite number of inequivalent representations
as a consequence of the nontrivial topology.
The algebra generated by quantum operators
is deformed by the central extension.
When the central extension is introduced,
it is shown that the zero-mode variables and the winding variables obey
a new commutation relation, which we call twist relation.
In addition, it is shown that the central extension makes momenta operators
obey anomalous commutators.
We demonstrate that topology enriches the structure of quantum field theories.
\end{minipage}
\end{center}
\newpage
\baselineskip 6mm
%
%
\section{Introduction}
As everyone recognizes,
quantum theory is established as an adequate and unique language
to describe microscopic phenomena
including condensed matter physics and also particle physics.
In this paper we would like to develop quantum theory
in the direction to seek for
another possibility which inheres in the theory itself.
 First we shall explain the well-known feature of quantum theory
and then we shall show the direction to pursue.
\par
To formulate a quantum theory
we start from commutation relations among generators.
Generators define an algebra, which is associative but not commutative.
When we have a classical theory,
we ordinarily bring commutators among canonical variables
by replacing the Poisson brackets.
Then we construct a Hilbert space
to be an irreducible representation space of the algebra.
An element of the algebra is chosen and called a hamiltonian.
Choice of a hamiltonian is restricted by physical requirements
like self-adjointness, positivity and symmetries.
The set formed by the algebra, the representation space and the hamiltonian
is called a quantum theory.
Then formulation itself is finished.
\par
Next tasks are to solve it;
we want to know eigenvalues of various observables,
especially we are interested in spectrum of the hamiltonian,
we want to know probability amplitude of transition
between an initial state and a final one.
 For physically interesting theories, those are difficult tasks.
Of course, we appreciate that they are worth hard work.
However it is not the direction
in which we would like to proceed in this paper.
We are interested in a rather formal aspect.
\par
A feature of quantum theory is that it requires a representation space;
operators must be provided with operands.
A commutator in quantum theory corresponds
to a Poisson bracket in classical theory.
But the Hilbert space in quantum theory
has no correspondence in classical theory.
State vectors, the superposition principle and amplitudes
are characteristic concepts of quantum theory.
We begin with an algebra of operators,
then we construct a representation space.
That is a unique procedure of quantum theory.
\par
Here arise two questions;
does a representation space exist?
Is it unique?
If there are inequivalent representations,
calculation based on a different space gives
a different answer for a physical quantity, for example, spectrum or amplitude.
\par
When we consider a particle in a Euclidean space,
we begin with the usual canonical commutation relations;
\begin{eqnarray}
	&&
	[ \, \hat{x}_i , \hat{x}_j \, ] = 0, \;\;\; ( i, j = 1, \cdots , n )
	\label{1.1}
	\\
	&&
	[ \, \hat{x}_i , \hat{p}_j \, ] = i \, \delta_{ij},
	\label{1.2}
	\\
	&&
	[ \, \hat{p}_i , \hat{p}_j \, ] = 0.
	\label{1.3}
\end{eqnarray}
According to von Neumann's theorem
the irreducible representation of the above algebra exists uniquely
within a unitary equivalence class.
Therefore there is no problem in choice of a Hilbert space.
Although one may use the wave function representation
and another may use the harmonic oscillator representation,
both obtain a same result for calculation of a physical quantity.
\par
Is there no need to worry about existence and uniqueness of a representation?
Actually it is needed.
We have encountered a situation in which the uniqueness is violated,
when we consider a quantum field theory.
In a quantum field theory
we construct a representation space
by defining a vacuum state and a Fock space.
It was found that in several models there exist inequivalent vacuum states
and they result in inequivalent Fock spaces.
The different vacua are characterized
by its transformation property under a certain symmetry operation.
Such a situation is called spontaneous symmetry breaking (SSB).
The discovery of SSB opened rich aspects of quantum field theories
and led to deep understanding of the nature.
\par
A field theory deals with a system which has infinite degrees of freedom,
namely it is defined with a infinite number of generators
which are called field variables.
It is known that SSB is related to the infinity of degrees of freedom.
On the other hand a particle has only finite degrees of freedom.
Is there no occurence of inequivalent representations
in a quantum theory of a particle?
(Usually a quantum theory with finite degrees of freedom
is called a quantum mechanics.)
\par
A strange result was found;
when Ohnuki and Kitakado~\cite{OK} investigated
a quantum mechanics of a particle on a circle $ S^1 $,
they showed that there are a infinite number of inequivalent Hilbert spaces.
Those spaces are parametrized by a continuous parameter $ \alpha $
ranging from 0 to 1.
What they have shown is that even a system with finite degrees of freedom
can possess inequivalent representations
when topology of the system is nontrivial.
After that work, they studied a quantum mechanics on a sphere
$ S^n ( n \ge 2 )$
and showed existence of an infinite number of inequivalent Hilbert spaces
specified by a discrete index.
\par
Let us turn to field theories.
The scalar field theory is a field-theoretical correspondence
to the quantum mechanics of a particle in a Euclidean space.
This field theory is quantized
by requiring the canonical commutators and constructing the Fock space.
There is also a correspondence in field theories
to a quantum mechanics on a nontrivial manifold.
It is a nonlinear sigma model because it has a manifold-valued field.
 For a review on nonlinear sigma models, see the reference~\cite{BKY}.
\par
Originally
the nonlinear sigma model is designed
to describe behavior of Nambu-Goldstone (NG) bosons at low energy scale.
NG bosons are massless excitations associated with SSB.
When a continuous symmetry specified by a group $ G $ is broken
to a smaller symmetry specified by a subgroup $ H $, vacua form a manifold
which is called a homogeneous space $ G/H $.
In this model NG bosons are described by a field taking values in $ G/H $.
It is already known~\cite{Mackey} that
even the quantum mechanics on $ G/H $ has inequivalent Hilbert spaces.
Therefore it is naturally expected that a quantum field theory with
a manifold-valued field may possess inequivalent Hilbert spaces.
However in a usual approach,
the nonlinear sigma model is quantized by the canonical quantization
as the scalar field theory and is solved by the perturbative method.
Thus the topological nature of the theory is missed
and only a Fock space provides a representation.
\par
If topological properties of field theories do not play an important role
in physical application,
we could neglect them.
However
we know several models which have nontrivial topology
and in which topology plays an important role.
 For instance,
the sine-Gordon model has topological kinks~\cite{soliton};
some gauge theories have topologically nontrivial vacua,
so-called $ \theta $-vacua~\cite{Jackiw};
some nonlinear sigma models have the Wess-Zumino-Witten term,
which reflects anomaly and topology of vacua~\cite{Witten};
the configuration space of nonabelian gauge fields modulo gauge transformations
has an extremely complicated topology
and causes the Gribov problem~\cite{Gribov},
and so on.
Hence we would like to develop quantum field theories
respecting topological nature.
\par
One of the aims of this paper is to demonstrate that
there are a lot of possibilities in constructing of quantum field theories,
even if they are identical as classical theories.
The second aim is to clarify
the relation between topology and quantization.
In the section 2 we will give a review of the quantum mechanics on $ S^1 $.
We will discuss physical implication of the existence
of inequivalent representations.
The section 3 is a main part of the present paper.
There we consider a simple but nontrivial field-theoretical model,
the abelian sigma model in (1+1) dimensions.
We propose a definition of an algebra,
which is quite different from the canonical one.
Then we will construct Hilbert spaces and classify inequivalent ones.
The section 4 is devoted to discussion
on the results and directions for future development.
This paper is a detailed and extended sequel to the previous work~\cite{sigma}.
\newpage
%
%
\section{Quantum mechanics on $ S^1 $}
Here we shall give a review of the quantum mechanics on $ S^1 $~\cite{OK}.
Idea in it is useful for the next step in the field theory.
More detailed consideration is found in the reference~\cite{S1}.
\subsection{Motivation}
Let us consider a particle moving on a circle $ S^1 $.
Its position is indicated by the angle coordinate $ \theta $.
The coordinate $ \theta + 2 \pi $ indicates the same point as $ \theta $ does.
Hence $ \theta $ is multivalued function on $ S^1 $.
Because of its nontrivial topology,
a continuous and single-valued coordinate does not exist on $ S^1 $.
\par
Now let us quantize it.
We should define an algebra.
If we take the canonical approach,
assume that $ \hat{\theta} $ and $ \hat{P} $ are self-adjoint operators
satisfying
\begin{equation}
	[ \, \hat{\theta} , \hat{P} \, ] = i.
	\label{2.1}
\end{equation}
Next we should construct a representation space.
If we define eigenstates of $ \hat{\theta} $ by
\begin{equation}
	\hat{\theta} \, | \theta \ket
	=
	\theta \, | \theta \ket,
	\label{2.2}
\end{equation}
we deduce that
\begin{equation}
	\bra \theta | \hat{P} | \psi \ket
	=
	- i \frac{\partial}{\partial \theta} \bra \theta | \psi \ket
	\label{2.3}
\end{equation}
for an arbitrary state $ | \psi \ket $.
Notice that the eigenvalue of $ \hat{\theta} $ ranges over
from $ - \infty $ to $ + \infty $;
it is {\it not true} that $ | \theta + 2 \pi \ket = | \theta \ket $.
What we have constructed is just the quantum mechanics on $ \R $, not $ S^1 $.
\par
What is wrong?
It is wrong to use the multivalued coordinate $ \hat{\theta} $
as a generator of the algebra.
We must use a well-defined single-valued generator from the beginning.
\par
As a substitute for $ \theta $, Ohnuki and Kitakado proposed to use
$ U = e^{ i \theta } $, which is complex-valued in the classical theory
and a unitary operator in the quantum theory.
 From (\ref{2.1}) we can deduce
\begin{equation}
	[ \, e^{ i \hat{\theta}} , \hat{P} \, ] = - e^{ i \hat{\theta}}.
	\label{2.4}
\end{equation}
But they did not regard $ \hat{\theta} $ as a generator.
Instead they took existence of a unitary operator $ \hat{U} $
and a self-adjoint operator $ \hat{P} $ satisfying
\begin{equation}
	[ \, \hat{U} , \hat{P} \, ] = - \hat{U}
	\label{2.5}
\end{equation}
as an assumption.
\subsection{Representations}
Now we see their construction of representations of
the algebra defined by the above commutation relation (\ref{2.5}).
Since $ \hat{P} $ is a self-adjoint operator,
it has an eigenvector with a real eigenvalue $ \alpha $;
\begin{equation}
	\hat{P} \, | \alpha \ket = \alpha \, | \alpha \ket,
	\qquad
	\bra \alpha | \alpha \ket = 1.
	\label{2.6}
\end{equation}
$ \hat{U} $ raises the eigenvalues of $ \hat{P} $;
\begin{eqnarray}
	\hat{P} \, \hat{U} | \alpha \ket
	&=&
	( \, [ \, \hat{P} , \hat{U} \, ] + \hat{U} \hat{P} ) \, | \alpha \ket
	\nonumber\\
	&=&
	( \, \hat{U} + \hat{U} \alpha ) \, | \alpha \ket
	\nonumber\\
	&=&
	( 1 + \alpha ) \hat{U} | \alpha \ket.
	\label{2.7}
\end{eqnarray}
Inversely, $ \hat{U}^{-1} = \hat{U}^{\dagger} $ lowers them. Then we define
\begin{equation}
	| n + \alpha \ket = \hat{U}^n | \alpha \ket,
	\quad
	( n = 0 , \pm 1 , \pm 2 , \cdots )
	\label{2.8}
\end{equation}
which have the following properties
\begin{eqnarray}
	&&
	\hat{P} \, | n + \alpha \ket = ( n + \alpha ) \, | n + \alpha \ket,
	\label{2.9}
	\\
	&&
	\bra m + \alpha | n + \alpha \ket = \delta_{mn}.
	\label{2.10}
\end{eqnarray}
The latter follows from self-adjointness of $ \hat{P} $
and unitarity of $ \hat{U} $.
With a fixed real number $ \alpha $, we define a Hilbert space
$ H_{\alpha} $ by completing the vector space of linear combinations of
$ | n + \alpha \ket $ ($ n $ : integer). Equation (\ref{2.9}) with
\begin{equation}
	\hat{U} \, | n + \alpha \ket = | n + 1 + \alpha \ket
	\label{2.11}
\end{equation}
defines an irreducible representation of the algebra (\ref{2.5})
on $ H_{\alpha} $.
\par
$ H_{\alpha} $ and $ H_{\beta} $ are unitary equivalent
if and only if the difference $ ( \alpha - \beta ) $ is an integer.
Therefore the classification of irreducible representations
of the algebra (\ref{2.5}) has been completed;
the whole of inequivalent irreducible representation spaces is
$ \{ H_{\alpha} \} \: ( 0 \leq \alpha < 1 ) $.
We call them Ohnuki-Kitakado representations.
\subsection{Physical implications}
\subsubsection{Wave function}
The physical meaning of the parameter $ \alpha $ seems obscure.
To clarify it
they~\cite{OK} studied eigenstates of the position operator $ \hat{U} $.
If we put
\begin{equation}
	| \theta \ket
	=
	\sum_{n \, = \, - \infty}^{\infty} \,
	e^{ - i n \theta } \, | n + \alpha \ket,
	\label{2.12}
\end{equation}
it follows that
\begin{eqnarray}
	&&
	\hat{U} \, | \theta \ket = e^{ i \theta } \, | \theta \ket,
	\label{2.13}
	\\
	&&
	| \theta + 2 \pi \ket = | \theta \ket,
	\label{2.14}
	\\
	&&
	\bra \theta | \theta' \ket =
	2 \pi \, \delta( \theta - \theta' ).
	\label{2.15}
\end{eqnarray}
In the last equation it is assumed that
the $ \delta $-function is periodic with periodicity $ 2 \pi $.
Eq. (\ref{2.14}) is a desired property for the quantum mechanics on $ S^1 $.
It is natural to call $ \hat{U} $ a position operator due to this property.
On the other hand, if we define $ \hat{V}(\mu) = \exp( -i \mu \hat{P} ) $
for a real number $ \mu $, (\ref{2.5}) implies that
\begin{equation}
	\hat{V}^\dagger (\mu) \, \hat{U} \, \hat{V}(\mu)
	= e^{ i \mu } \, \hat{U}.
	\label{2.16}
\end{equation}
Thus
$ \hat{U} \, \hat{V}(\mu) | \theta \ket
= e^{ i ( \theta + \mu ) } \, \hat{V}(\mu) | \theta \ket $
is an immediate consequence.
A direct calculation shows that
\begin{equation}
	\hat{V}( \mu ) | \theta \ket
	=
	e^{ -i \alpha \mu } \, | \theta + \mu \ket,
	\label{2.17}
\end{equation}
which says that $ \hat{P} $ is a generator of translation along $ S^1 $.
It should be noticed that an extra phase factor $ e^{ - i \alpha \mu } $
is multiplied.
These states
$ | \theta \ket \, ( 0 \le \theta < 2 \pi ) $
define a wave function $ \psi ( \theta ) $
for an arbitrary state $ | \psi \ket \in H_\alpha $
by $ \psi ( \theta ) = \bra \theta | \psi \ket $.
This definition gives an isomorphism between $ H_\alpha $
and $ L^2 ( S^1 ) $ that is a space of square-integrable functions on $ S^1 $.
A bit calculation shows that the operators act on the wave function as
\begin{eqnarray}
	&&
	\hat{U} \psi ( \theta )
	= \bra \theta | \hat{U} | \psi \ket
	= e^{ i \theta } \, \psi ( \theta ),
	\label{2.18}
	\\
	&&
	\hat{V}(\mu) \psi ( \theta )
	= \bra \theta | \hat{V}(\mu) | \psi \ket
	= e^{ -i \alpha \mu } \, \psi ( \theta - \mu ),
	\label{2.19}
	\\
	&&
	\hat{P} \psi ( \theta )
	= \bra \theta | \hat{P} | \psi \ket
	= \Bigl( - i \frac{\partial}{\partial \theta} + \alpha \Bigr)
	\psi ( \theta ).
	\label{2.20}
\end{eqnarray}
In the last expression
the parameter $ \alpha $ looks like the vector potential
for magnetic flux $ \Phi = 2 \pi \alpha $ surrounded by $ S^1 $.
It should be noticed that
$ \alpha $ cannot be removed by gauge transformation
$ \psi(\theta) \to \psi'(\theta) = \kappa(\theta) \, \psi(\theta) $,
where $ \kappa $ is a function from $ S^1 $ to $ U(1) $.
If we could chose $ \kappa(\theta) = e^{ i \alpha \theta } $,
\begin{equation}
	\Bigl( - i \frac{\partial}{\partial \theta} + \alpha \Bigr)
	\psi ( \theta )
	=
	e^{ -i \alpha \theta }
	\Bigl( - i \frac{\partial}{\partial \theta} \Bigr)
	\psi' ( \theta )
	\label{2.21}
\end{equation}
thus $ \alpha $ disappeared.
However, because $ \psi'(2\pi) = e^{ i \alpha 2 \pi } \psi'(0) $,
$ \psi' $ is not a periodic function.
Hence $ \psi' $ does not remain in $ L^2(S^1) $.
In picture of wave functions,
it is the boundary condition $ \psi(2\pi) = \psi(0) $
what obstructs elimination of $ \alpha $
and therefore causes inequivalent representations.
In the quantum mechanics on $ \R $, there is no such boundary condition,
thus such an extra term can be wiped away by gauge transformation.
\subsubsection{Spectrum}
To see a physical effect of the parameter $ \alpha $
let us consider a free particle on $ S^1 $.
A free particle is defined by the hamiltonian
\begin{equation}
	\hat{H} = \frac12 \hat{P}^2.
	\label{2.22}
\end{equation}
Its eigenvalue problem is trivially solved by
\begin{equation}
	\hat{H} \, | n + \alpha \ket
	=
	\frac12 ( n + \alpha )^2 | n + \alpha \ket.
	\label{2.23}
\end{equation}
Apparently, the spectrum depends on the parameter $ \alpha $.
 For $ \alpha = m $ ($ m $ : integer),
all the eigenvalues but one of the ground state are doubly degenerate.
While for $ \alpha = m + \frac 12 $,
the all eigenvalues are doubly degenerate.
 For other values of $ \alpha $, there is no degeneracy.
It is shown in the previous work~\cite{S1} that
these degeneracies reflect the parity symmetry.
As $ n+\alpha = (n-1)+(\alpha+1) $,
the spectrum on the Hilbert space $ H_\alpha $ is same as that on
$ H_{\alpha + 1} $.
Moreover, as $ (n+\alpha)^2 = (-n-\alpha )^2 $,
the spectrum of on $ H_\alpha $ is same to that on $ H_{- \alpha} $, too.
Therefore distinguishable values of $ \alpha $
range over $ 0 \le \alpha \le \frac 12 $.
%
%
\subsubsection{Path integral}
As another example to show a physical effect of the parameter $ \alpha $,
we shall see a path integral expression of the quantum mechanics on $ S^1 $.
We borrow a result from~\cite{path}.
When the hamiltonian is of the form
\begin{equation}
	\hat{H} = \frac 12 \, \hat{P}^{2} + V ( \hat{U}, \hat{U}^\dagger ),
	\label{2.24}
\end{equation}
they derived a path integral expression of transition amplitude as
\begin{eqnarray}
	K(\theta' , \theta ; t)
	&= &
	\bra \theta' | \exp( - i \hat{H} t ) | \theta \ket
	\nonumber
	\\
	& = &
	\sum_{n = -\infty}^{\infty}
	\int_{\mbox{\scriptsize winding} \: n \: \mbox{\scriptsize times}}
	{\cal D} \theta
	\, \exp( i S_{\mbox{\scriptsize eff}} ) ,
	\label{2.25}
\end{eqnarray}
where the effective action is defined as
\begin{equation}
	S_{\mbox{\scriptsize eff}}
	=
	\int d t
	\left[
		\frac 12 \biggl( \frac{d \theta}{d t} \biggr)^2
		- V ( \theta ) - \alpha \frac{d \theta}{d t}
	\right].
	\label{2.26}
\end{equation}
In (\ref{2.25}),
the integration is performed over paths winding $ n $ times around $ S^1 $ and
the summation is performed with respect to the winding number.
We would like to emphasize that the above path integral expression is derived
from the operator formalism.
It should be noticed that the global property---winding number---appears
from the operator formalism alone.
\\
%
%
\par
The last term in (\ref{2.26}), $ \int \alpha \, d \theta $ has no influence
on the equation of motion but has an observable effect on the amplitude.
To see the role of this term we rewrite (\ref{2.25}) as
\begin{equation}
	K(\theta' , \theta ; t)
	=
	e^{ - i \alpha (\theta' - \theta) }
	\sum_{n = -\infty}^{\infty}
	e^{ - i \alpha 2 \pi n }
	\int_{\mbox{\scriptsize winding} \: n \: \mbox{\scriptsize times}}
	{\cal D} \theta
	\, \exp( i S_0 ),
	\label{2.27}
\end{equation}
where $ S_0 = \int d t \, [ \frac 12 \dot{\theta}^2 - V(\theta) ] $.
An amplitude for a path winding $ n $ times is weighted by
the phase factor $ \omega_n = \exp( - i \alpha 2 \pi n ) $.
This phase factor causes observable interference effect;
this phenomenon is analogous to the Aharonov-Bohm effect.
 Furthermore, $ \omega_n $'s obey composition rule;
$ \omega_m \, \omega_n = \omega_{m+n} $,
which says that $ m $-times winding followed by $ n $-times one is equal
to $ (m+n) $-times one.
According to~\cite{Schulman},
$ \omega_n $ can be interpreted as a unitary representation
of the first homotopy group $ \pi^1(S^1) $.
\par
We conclude this section by repeating what has been shown.
To formulate the quantum mechanics on $ S^1 $
we should choose suitable generators to define the algebra.
We recognize that $ \theta $ is not suitable but $ U $ is suitable.
The algebra is defined by the commutation relation (\ref{2.5}).
Representation spaces are constructed
and inequivalent ones are parametrized by a continuous parameter
$ \alpha $ $ ( 0 \le \alpha < 1 ) $.
Inequivalent ones give different solutions to physical problems.
The role of $ \alpha $ resembles that of the vector potential.
Topology of $ S^1 $ is an obstruction
against elimination of such a vector potential.
In path integral picture,
$ \alpha $ characterizes homotopy of a path of the particle on $ S^1 $.
Accordingly, existence of $ \alpha $ reflects topology of $ S^1 $.
\newpage
%
%
\section{Abelian sigma model}
Here we shall consider the abelian sigma model as a generalization
of the quantum mechanics on $ S^1 $.
The abelian sigma model has a field which takes values in $ S^1 $.
This model is designed to describe an NG boson
associated with spontaneous breaking of $ U(1) $ symmetry.
We shall define two classes of algebras of the field theory;
the first one is a natural generalization of the quantum mechanics on $ S^1 $,
which is called an algebra without central extension;
the second one is its nontrivial generalization,
which is called an algebra with central extension.
 For both classes we shall construct representation spaces
combining the usual Fock representation
with the Ohnuki-Kitakado representation.
Topology of the model is carefully treated during the construction.
\subsection{Algebra}
\subsubsection{Definition}
To motivate definition of the algebra we will take three steps.
 First we start from the quantum mechanics on a Euclidean space $ \R^n $.
We shall give another expression to the canonical commutation relations
(\ref{1.1})-(\ref{1.3}).
If we put $ \hat{V}(a) = \exp ( - i \sum_j \, a_j \hat{p}_j ) $
for real numbers $ a = ( a_1 , \cdots , a_n ) $,
$ \hat{V}(a) $ is a unitary operator and satisfies
\begin{eqnarray}
	&&
	\hat{x}_j \, \hat{x}_k = \hat{x}_k \, \hat{x}_j,
	\label{3.1}
	\\
	&&
	\hat{V}^\dagger (a) \, \hat{x}_j \, \hat{V}(a) = \hat{x}_j + a_j,
	\label{3.2}
	\\
	&&
	\hat{V}(a) \, \hat{V}(b) = \hat{V}( a + b ).
	\label{3.3}
\end{eqnarray}
Geometrical meaning of the above algebra is obvious.
(\ref{3.1}) says that
coordinates $ \hat{x} $'s of configuration are simultaneously measurable.
(\ref{3.2}) implies that
configuration is movable by the displacement operator $ \hat{V}(a) $.
(\ref{3.3}) says that
displacement operators satisfy the composition law.
An irreducible representation is uniquely given by $ L^2 ( \R^n ) $.
\par
Second we turn to the scalar field theory in (1+1) dimensions.
The usual canonical commutation relations are
\begin{eqnarray}
	&&
	[ \, \hat{\varphi} (\sigma), \hat{\varphi} (\sigma') \, ] = 0,
	\;\;\; ( - \infty < \sigma, \sigma' < + \infty )
	\label{3.4}
	\\
	&&
	[ \, \hat{\varphi} (\sigma), \hat{\pi} (\sigma') \, ]
	= i \, \delta ( \sigma - \sigma' ),
	\label{3.5}
	\\
	&&
	[ \, \hat{\pi} (\sigma), \hat{\pi} (\sigma') \, ] = 0,
	\label{3.6}
\end{eqnarray}
where $ \sigma $ is a coordinate of the space.
$ \hat{\varphi} $ and $ \hat{\pi} $ are
distributions valued in hermite operators.
We introduce a unitary operator
\begin{equation}
	\hat{V} ( f )
	=
	\exp
	\left[
		- i \int_{-\infty}^{\infty}
		f(\sigma) \, \hat{\pi} (\sigma)
		\, d \sigma
	\right]
	\label{3.7}
\end{equation}
for a real-valued test function $ f $.
Then they satisfy
\begin{eqnarray}
	&&
	\hat{\varphi}(\sigma)  \, \hat{\varphi}(\sigma') =
	\hat{\varphi}(\sigma') \, \hat{\varphi}(\sigma),
	\label{3.8}
	\\
	&&
	\hat{V}^\dagger (f) \, \hat{\varphi}(\sigma) \, \hat{V}(f) =
	\hat{\varphi}(\sigma) + f(\sigma),
	\label{3.9}
	\\
	&&
	\hat{V}(f) \, \hat{V}(g) = \hat{V}(f+g),
	\label{3.10}
\end{eqnarray}
where $ f $ and $ g $ are arbitrary real-valued test functions.
The above algebra is interpreted as follows.
(\ref{3.8}) means that
the field configurations at separated points are simultaneously measurable.
(\ref{3.9}) implies that
a field configuration is movable arbitrarily by displacement operators.
(\ref{3.10}) is nothing but the composition property of displacements.
A representation is usually given in terms of the Fock space.
\par
As the third step we generalize the quantum mechanics on $ S^1 $
to the one on $ n $-dimensional torus $ T^{\, n} = ( S^1 )^n $.
We introduce unitary operators $ \hat{U}_j $ and
self-adjoint operators $ \hat{P}_j ( j = 1 , \cdots , n ) $.
Put $ \hat{V}( \mu ) = \exp( -i \sum_j \, \mu_j \hat{P}_j ) $
for $ \mu = ( \mu_1 , \cdots , \mu_n ) \in \R^n $.
Naive generalization of (\ref{2.5}) or (\ref{2.16}) leads
the following relations
\begin{eqnarray}
	&&
	\hat{U}_j \, \hat{U}_k = \hat{U}_k \, \hat{U}_j,
	\label{3.11}
	\\
	&&
	\hat{V}^\dagger (\mu) \, \hat{U}_j \, \hat{V}(\mu)
	= e^{ i \mu_j } \, \hat{U}_j,
	\label{3.12}
	\\
	&&
	\hat{V}(\mu) \, \hat{V}(\nu) = \hat{V}( \mu + \nu ).
	\label{3.13}
\end{eqnarray}
Geometrical meaning is so obvious that explanation is not repeated.
We only point out that
(\ref{3.12}) expresses action of $ \R^n $ on $ T^{\, n} $ by displacement.
Representations of this algebra are constructed
by tensor products of Ohnuki-Kitakado representations
$ H_{\alpha_1} \otimes \cdots \otimes H_{\alpha_n} $.
Therefore irreducible representations are parametrized by $ n $-tuple parameter
$ \alpha = ( \alpha_1 , \cdots , \alpha_n ) $.
\par
 Finally we turn to the abelian sigma model in (1+1) dimensions.
The space-time is assumed to be $ S^1 \times \R $.
On an equal-time space-slice,
the classical field variable is a map from $ S^1 $ to $ S^1 $.
So the configuration space of the model is $ Q = \map(S^1; S^1) $.
On the other hand
$ \Gamma = \map(S^1; U(1)) $ becomes a group by pointwise multiplication.
The group $ U(1) $ acts on $ S^1 $ by displacement.
Thus the group $ \Gamma $ acts on the configuration space $ Q $
by pointwise action,
that is to say,
for $ \gamma \in \Gamma $ and $ \phi \in Q $
let us define $ \gamma \cdot \phi \in Q $ by
\begin{equation}
	( \gamma \cdot \phi ) ( \sigma ) = \gamma(\sigma) \cdot \phi(\sigma)
	\;\;\;
	(\sigma \in S^1),
	\label{3.14}
\end{equation}
where $ \sigma $ denotes a point of the base space.
In the right-hand side
the multiplication indicates the action of $ U(1) $ on $ S^1 $.
\par
To clarify geometry of the classical theory,
we shall decompose the degrees of freedom of $ \phi \in Q $ and
$ \gamma \in \Gamma $.
In the classical sense we may rewrite $ \phi : S^1 \to S^1 \cong U(1) $ by
\begin{equation}
	\phi(\sigma) = U \, e^{ i \, ( N \sigma + \varphi(\sigma) ) },
	\label{3.15}
\end{equation}
where $ U \in U(1) $, $ N \in \Z $.
$ \varphi $ satisfies the no zero-mode condition;
\begin{equation}
	\map_0 (S^1; \R) =
	\{
	\varphi : S^1 \to \R
	\, | \,
	C^\infty , \,
	\int_0^{2 \pi} \varphi(\sigma) \, d \sigma = 0
	\}.
	\label{3.16}
\end{equation}
The decomposition (\ref{3.15}) says that
$ Q \cong S^1 \times \Z \times \map_0 ( S^1; \R ) $.
Geometrical meaning of this decomposition is apparent;
$ U $ describes the zero-mode or collective motion of the field $ \phi $;
$ N $ is nothing but the winding number;
$ \varphi $ describes fluctuation or local degrees of freedom of $ \phi $.
Topologically nontrivial parts are $ U $ and $ N $.
%
%
\par
Similarly $ \gamma : S^1 \to U(1) $ is also rewritten as
\begin{equation}
	\gamma ( \sigma ) = e^{ i ( \mu + m \sigma + f(\sigma) ) },
	\label{3.17}
\end{equation}
where $ \mu \in \R $, $ m \in \Z $ and $ f \in \map_0 ( S^1 ; \R ) $.
The action (\ref{3.14}) of $ \gamma $ on $ \phi $ is decomposed into
\begin{eqnarray}
	U & \to & e^{ i \mu } \, U,
	\label{3.18}
	\\
	N & \to & N + m,
	\label{3.19}
	\\
	\varphi(\sigma) & \to & \varphi(\sigma) + f(\sigma),
	\label{3.20}
\end{eqnarray}
according to (\ref{3.15}) and (\ref{3.17}).
Thus the first component of $ \gamma $ (\ref{3.17}) translates the zero-mode;
the second one changes the winding number;
the third one gives a homotopic deformation.
\par
To quantize this system let us assume that
$ \hat{\phi}(\sigma) $ is a unitary operator for each point $ \sigma \in S^1 $
and
$ \hat{V}(\gamma) $ is a unitary operator for each element
$ \gamma \in \Gamma $.
Moreover we define an algebra generated by $ \hat{\phi}(\sigma) $ and
$ \hat{V}(\gamma) $ with the following relations
\begin{eqnarray}
	&&
	\hat{\phi}(\sigma)  \, \hat{\phi}(\sigma') =
	\hat{\phi}(\sigma') \, \hat{\phi}(\sigma),
	\;\;\;
	( \sigma, \, \sigma' \in S^1 )
	\label{3.21}
	\\
	&&
	\hat{V}^\dagger(\gamma) \, \hat{\phi}(\sigma) \, \hat{V}(\gamma) =
	\gamma(\sigma) \, \hat{\phi}(\sigma),
	\label{3.22}
	\\
	&&
	\hat{V}(\gamma_1) \, \hat{V}(\gamma_2) =
	e^{ - i c ( \gamma_1, \gamma_2 ) } \,  \hat{V}(\gamma_1 \gamma_2)
	\;\;\;
	( \gamma_1, \, \gamma_2 \in \Gamma ).
	\label{3.23}
\end{eqnarray}
At the last line a function $ c : \Gamma \times \Gamma \to \R $
is called a central extension, which satisfies the cocycle condition
\begin{equation}
	  c( \gamma_1 , \gamma_2 )
	+ c( \gamma_1 \gamma_2 , \gamma_3 )
	= c( \gamma_1 , \gamma_2 \gamma_3 )
	+ c( \gamma_2 , \gamma_3 )
	\;\;\;
	( \mbox{mod} \: 2 \pi )
	\label{3.24}
\end{equation}
to ensure associativity
$ ( \hat{V}(\gamma_1)   \hat{V}(\gamma_2) ) \hat{V}(\gamma_3)
=   \hat{V}(\gamma_1) ( \hat{V}(\gamma_2)   \hat{V}(\gamma_3) ) $.
If $ c \equiv 0 $, the algebra (\ref{3.21})-(\ref{3.23}) is
a straightforward generalization of (\ref{3.11})-(\ref{3.13})
to a system with infinite degrees of freedom.
We call the algebra defined by (\ref{3.21})-(\ref{3.23})
the fundamental algebra of the abelian sigma model.
\par
We should explain
why such an extra phase factor $ e^{- ic(\gamma_1, \gamma_2) } $
is introduced.
$ \hat{V}(\gamma) $ acts on $ \hat{\phi}(\sigma) $
by adjoint action as shown in (\ref{3.22}).
This action expresses the action of the group $ \Gamma $ on $ Q $
in terms of quantum operators.
It satisfies the composition law
\begin{equation}
	\hat{V}^\dagger(\gamma_2) \, \hat{V}^\dagger(\gamma_1) \,
	\hat{\phi}(\sigma) \,
	\hat{V}(\gamma_1) \hat{V}(\gamma_2)
	=
	\hat{V}^\dagger(\gamma_1 \gamma_2) \,
	\hat{\phi}(\sigma) \,
	\hat{V}(\gamma_1 \gamma_2),
	\label{3.25}
\end{equation}
which is demanded from geometrical viewpoint.
However the above composition law does {\it not} imply that
$ \hat{V}(\gamma_1) \, \hat{V}(\gamma_2) = \hat{V}(\gamma_1 \gamma_2) $.
In other words, $ \hat{V} $ is not necessarily
a genuine unitary representation of the group $ \Gamma $.
We always have a possibility to insert a phase factor as done in (\ref{3.23}).
In other words, $ \hat{V} $ may be a projective unitary representation.
If we can find a function $ b : \Gamma \to \R $ such that
\begin{equation}
	c( \gamma_1, \gamma_2 )
	=
	b( \gamma_1 ) + b( \gamma_2 ) - b( \gamma_1 \gamma_2 ),
	\;\;\;
	( \mbox{mod} \: 2 \pi )
	\label{3.26}
\end{equation}
by defining $ \tilde{V} (\gamma) = e^{ i b (\gamma) } \hat{V} (\gamma) $,
(\ref{3.23}) results in a genuine unitary representation
$ \tilde{V}(\gamma_1) \, \tilde{V}(\gamma_2) = \tilde{V}(\gamma_1 \gamma_2) $.
In that case $ c $ is called a coboundary of $ b $
and denoted by $ c = \delta b $.
A coboundary $ c = \delta b $ identically satisfies
the condition (\ref{3.24}),
namely a coboundary is always a cocycle.
We usually demand both $ b $ and $ c $ to be continuous functions
from $ \Gamma $ to $ \R / 2\pi\Z $,
since $ \Gamma $ is continuous.
A class of cocycles modulo coboundaries is called a cohomology.
Existence of a nontrivial cohomology depends on
topology of the group $ \Gamma $.
The group considered now is a so-called loop group
$ \Gamma = \map(S^1; G) $ with $ G=U(1) $.
Its nontrivial topology allows for existence of a nontrivial cohomology.
On the other hand, when $ \Gamma = \R^n $, all cocycles are coboundaries.
Thus there is no need to insert such a central extension into (\ref{3.3})
when we considered the quantum mechanics on $ \R^n $.
\par
We add a comment.
In quantum field theories, we have met such extensions of algebras
when we study anomalous gauge theories and conformal field theories.
In a gauge theory in (3+1) dimensions with a gauge group $ G $,
gauge transformations form a group $ \Gamma = \map(S^3, G) $.
Anomaly is deeply related to topology of $ \Gamma $.
When a nontrivial cohomology exists, extra terms are added
to the commutators of the Gauss law constraints~\cite{anomaly}.
Similarly in two dimensional conformal field theories~\cite{Goddard},
the algebra of energy-momentum tensor is deformed by a central extension
due to the conformal anomaly
and becomes the Virasoro algebra.
In the two dimensional Wess-Zumino-Witten (WZW) model~\cite{Witten},
the current algebra is also deformed by a central extension
and becomes the Kac-Moody algebra.
Its deformation is caused by the WZW term brought into the lagrangian
of a nonlinear sigma model.
A necessary condition for existence of the WZW term is
that $ G $ has nontrivial cohomology $ H^3(G) $.
But $ U(1) $ does not satisfy it.
We do not yet know a physical reason why we must introduce
a central extension into the abelian sigma model.
Even a reason to chose a specific central extension is not clear.
In the case of the WZW model,
QCD as an underlying theory tells the reason to bring the anomaly.
But the physical meaning of the central extension in our model
still remains obscure.
\par
Now let us return to the fundamental algebra (\ref{3.21})-(\ref{3.23}).
As a nontrivial central extension for $ \gamma_1 $ given by (\ref{3.17}) and
\begin{equation}
	\gamma_2 ( \sigma ) = e^{ i ( \nu + n \sigma + g(\sigma) ) },
	\label{3.27}
\end{equation}
we define
\begin{equation}
	c( \gamma_1 , \gamma_2 )
	=
	k
	\left\{
		m \nu - n \mu
		+
		\frac{1}{4 \pi} \int_0^{2 \pi}
		\biggl(
			  \frac{d f}{d \sigma} g(\sigma)
			- f(\sigma) \frac{d g}{d \sigma}
		\biggr)
		d \sigma
	\right\},
	\label{3.28}
\end{equation}
where $ k $ is an integer.
This central extension is the simplest but nontrivial one
which is invariant under the action of the group of orientation-preserving
diffeomorphims $ \diff( S^1 ) $;
$ c( \gamma_1 \circ \omega , \gamma_2 \circ \omega )
= c( \gamma_1 , \gamma_2 ) $
for any $ \omega \in \diff( S^1 ) $.
The group $ \Gamma $ associated with such an invariant central extension
is called a Kac-Moody group of rank $ k $.
The relation (\ref{3.23}) says that
$ \hat{V} $ is a unitary representation of the Kac-Moody group.
 For classification of central extensions see the literature~\cite{Segal}.
\subsubsection{Algebra without central extension}
According to decomposition of classical variables
(\ref{3.15}) and (\ref{3.17}), quantum operators are also to be decomposed.
 For simplicity we consider the fundamental algebra (\ref{3.21})-(\ref{3.23})
without the central extension, that is, here we restrict $ c \equiv 0 $.
\par
Corresponding to (\ref{3.15}), we introduce
a unitary operator $ \hat{U} $,
a self-adjoint operator $ \hat{N} $ satisfying
\begin{equation}
	\exp( 2 \pi i \hat{N} ) = 1,
	\label{3.29}
\end{equation}
which is called the integer condition for $ \hat{N} $,
and
a distribution $ \hat{\varphi} ( \sigma ) $ valued in hermite operators
and constrained by
\begin{equation}
	\int_0^{ 2 \pi } \hat{\varphi} ( \sigma ) \, d \sigma = 0.
	\label{3.30}
\end{equation}
We demand that the quantum field $ \hat{\phi} ( \sigma ) $ is expressed
in terms of these as
\begin{equation}
	\hat{\phi} ( \sigma )
	=
	\hat{U} \, e^{ i \, ( \hat{N} \sigma + \hat{\varphi} (\sigma) ) }.
	\label{3.31}
\end{equation}
Actually this equation is too naive.
Because $ \hat{\varphi} (\sigma) $ is an operator-valued distribution,
its exponentiation $ \exp( i \, \hat{\varphi} ( \sigma ) ) $ is ill-defined.
To make it well-defined we should regularize its divergence.
This problem is postponed until discussion on the normal ordering.
\par
Next, corresponding to (\ref{3.17}) we introduce
a self-adjoint operator $ \hat{P} $,
a unitary operator $ \hat{W} $,
and
a distribution $ \hat{\pi} ( \sigma ) $ valued in hermite operators
and constrained by
\begin{equation}
	\int_0^{ 2 \pi } \hat{\pi} ( \sigma ) \, d \sigma = 0,
	\label{3.32}
\end{equation}
When $ \gamma $ is given by (\ref{3.17}),
the operator $ \hat{V} ( \gamma ) $ is defined by
\begin{equation}
	\hat{V} ( \gamma )
	=
	e^{ - i \mu \hat{P} } \,
	\hat{W}^m
	\exp
	\left[
	- i \int_0^{2 \pi} f( \sigma ) \, \hat{\pi} ( \sigma )
	\, d \sigma
	\right].
	\label{3.33}
\end{equation}
\par
Using these operators the relation (\ref{3.22}) is now rewritten as
\begin{eqnarray}
	&&
	e^{ i \mu \hat{P} } \, \hat{U} \, e^{ - i \mu \hat{P} }
	=
	e^{ i \mu } \, \hat{U},
	\label{3.34}
	\\
	&&
	\hat{W}^\dagger \, \hat{N} \, \hat{W} = \hat{N} + 1,
	\label{3.35}
	\\
	&&
	\exp
	\left[
		  i \int_0^{2 \pi} f( \sigma ) \hat{\pi} ( \sigma ) d \sigma
	\right]
	\hat{\varphi} ( \sigma )
	\exp
	\left[
		- i \int_0^{2 \pi} f( \sigma ) \hat{\pi} ( \sigma ) d \sigma
	\right]
	=
	\hat{\varphi} ( \sigma ) + f ( \sigma ).
	\label{3.36}
\end{eqnarray}
These represent the action of $ \Gamma $ on $Q$ by (\ref{3.18})-(\ref{3.20}).
Observing the relation (\ref{3.35}), we call $ \hat{N} $ and $ \hat{W} $
the winding number and the winding operator, respectively.
They are also rewritten in terms of commutation relations
\begin{eqnarray}
	&&
	[ \, \hat{P} , \hat{U} \, ] = \hat{U},
	\label{3.37}
	\\
	&&
	[ \, \hat{N} , \hat{W} \, ] = \hat{W},
	\label{3.38}
	\\
	&&
	[ \, \hat{\varphi} ( \sigma ) , \hat{\pi} ( \sigma' ) \, ]
	=
	i \Bigl( \delta( \sigma - \sigma' ) - \frac{1}{2 \pi} \Bigr),
	\label{3.39}
\end{eqnarray}
with all other vanishing commutators.
In (\ref{3.39}) it is understood that the $ \delta $-function is defined
on $ S^1 $.
\subsubsection{Algebra with central extension}
Before constructing representations, we reexpress the fundamental algebra
with the central extension (\ref{3.28})
respecting the decomposition (\ref{3.15}) and (\ref{3.17}).
The decomposition (\ref{3.31}) of $ \hat{\phi} $ does not need to be changed.
On the other hand the decomposition (\ref{3.33}) of $ \hat{V} $
should be modified a little.
We formally introduce an operator $ \hat{\Omega} $ by
\begin{equation}
	\hat{W} = e^{ - i \, \hat{\Omega} }.
	\label{3.40}
\end{equation}
Although $ \hat{W} $ itself is well-defined,
$ \hat{\Omega} $ is ill-defined.
If $ \hat{\Omega} $ exists, (\ref{3.38}) would imply
$ [ \, \hat{N} , \hat{\Omega} \, ] = i $, 
which is nothing but the canonical commutation relation.
Therefore $ \hat{N} $ should have a continuous spectrum,
that contradicts the integer condition (\ref{3.29}).
Consequently $ \hat{\Omega} $ must be eliminated after calculation.
Bearing the above remark in mind, we replace (\ref{3.33}) by
\begin{equation}
	\hat{V} ( \gamma )
	=
	\exp
	\left[
		- i
		\Bigl(
			\mu \hat{P}
			+
			m \, \hat{\Omega}
			+
			\int_0^{2 \pi} f( \sigma ) \, \hat{\pi}(\sigma )
			\, d \sigma
		\Bigr)
	\right].
	\label{3.41}
\end{equation}
 For the central extension (\ref{3.28}) of rank $ k $,
addition of the following commutation relations to (\ref{3.37})-(\ref{3.39})
is enough to satisfy the fundamental algebra;
\begin{eqnarray}
	&&
	[ \, \hat{P} , \hat{\Omega} \, ] = - 2 i k,
	\label{3.42}
	\\
	&&
	[ \, \hat{\pi} (\sigma) , \hat{\pi} (\sigma') \, ]
	=
	- \, \frac{i k}{\pi} \, \delta' ( \sigma - \sigma' ).
	\label{3.43}
\end{eqnarray}
We assume that all other commutators vanish.
\par
Below we verify that they are sufficient for the fundamental algebra.
A useful formula is
\begin{equation}
	e^{\hat{X}} \, e^{\hat{Y}}
	=
	e^{\frac12 [ \, \hat{X}, \hat{Y} \, ]} \, e^{ \hat{X} + \hat{Y} },
	\label{3.44}
\end{equation}
which is valid when
$ [ \, \hat{X}, [ \, \hat{X}, \hat{Y} \, ] \, ]
= [ \, \hat{Y}, [ \, \hat{X}, \hat{Y} \, ] \, ]
= 0 $.
This formula yields
\begin{equation}
	e^{ - i ( \mu \hat{P} + m \, \hat{\Omega} ) } \,
	e^{ - i ( \nu \hat{P} + n \, \hat{\Omega} ) }
	=
	e^{ - \frac12 ( m \nu - n \mu ) [ \, \hat{\Omega}, \hat{P} \, ] } \,
	e^{ - i ( ( \mu + \nu ) \hat{P} + ( m + n ) \hat{\Omega} ) }.
	\label{3.45}
\end{equation}
If we substitute (\ref{3.42}), we obtain
\begin{equation}
	e^{ - i ( \mu \hat{P} + m \, \hat{\Omega} ) } \,
	e^{ - i ( \nu \hat{P} + n \, \hat{\Omega} ) }
	=
	e^{ - i k ( m \nu - n \mu ) } \,
	e^{ - i ( ( \mu + \nu ) \hat{P} + ( m + n ) \hat{\Omega} ) },
	\label{3.46}
\end{equation}
which is coincident with (\ref{3.23}) for the central extension (\ref{3.28}).
Similarly the formula (\ref{3.44}) implies
\begin{eqnarray}
	&&
	e^{ - i \int_0^{2 \pi} f( \sigma ) \hat{\pi}(\sigma ) d \sigma } \,
	e^{ - i \int_0^{2 \pi} g( \sigma') \hat{\pi}(\sigma') d \sigma'}
	\nonumber\\
	=
	&&
	e^{ - \frac12
	\int_0^{2 \pi} \int_0^{2 \pi} f( \sigma ) g( \sigma')
	[ \, \hat{\pi}(\sigma ), \hat{\pi}(\sigma') \, ]
	d \sigma \, d \sigma' }
	\,
	e^{ - i \int_0^{2 \pi} ( f( \sigma ) + g ( \sigma ) )
	\hat{\pi}(\sigma ) d \sigma }.
	\label{3.47}
\end{eqnarray}
Substitution of (\ref{3.43}) yields
\begin{eqnarray}
	&&
	- \frac12
	\int_0^{2 \pi} \int_0^{2 \pi} f( \sigma ) g( \sigma')
	[ \, \hat{\pi}(\sigma ), \hat{\pi}(\sigma') \, ]
	d \sigma \, d \sigma'
	\nonumber
	\\
	= &&
	\frac{i k}{2 \pi}
	\int_0^{2 \pi} \int_0^{2 \pi} f( \sigma ) g( \sigma')
	\, \delta' ( \sigma - \sigma' )
	d \sigma \, d \sigma'
	\nonumber
	\\
	= &&
	\frac{i k}{4 \pi}
	\int_0^{2 \pi}
	( - f'( \sigma ) g( \sigma) + f( \sigma ) g'( \sigma) )
	d \sigma.
	\label{3.48}
\end{eqnarray}
Hence (\ref{3.47}) coincides with (\ref{3.23})
for the central extension (\ref{3.28}).
\par
Using (\ref{3.40}) Eq. (\ref{3.42}) implies
\begin{equation}
	[ \, \hat{P} , \hat{W} \, ] = - 2 k \, \hat{W},
	\label{3.49}
\end{equation}
which says that the zero-mode momentum $ \hat{P} $ is decreased
by $ 2 k $ units when the winding number $ \hat{N} $ is increased by one unit
under the operation of $ \hat{W} $.
This is an inevitable consequence of the central extension.
We call this phenomenon ``twist''.
Using (\ref{3.42}) and (\ref{3.44}), the decomposition (\ref{3.41}) results in
\begin{equation}
	\hat{V} ( \gamma )
	=
	e^{ - i k m \mu }
	\,
	\exp
	\left[
		- i
		\Bigl(
			\mu \hat{P}
			+
			\int_0^{2 \pi} f( \sigma ) \, \hat{\pi}(\sigma )
			\, d \sigma
		\Bigr)
	\right]
	\hat{W}^m.
	\label{3.50}
\end{equation}
\par
Here we summarize a temporal result.
Generators of the fundamental algebra are decomposed
as (\ref{3.31}) and (\ref{3.50})
considering topological nature of the model.
They are constrained by the no-zero-mode condition (\ref{3.30}), (\ref{3.32})
and the integer condition (\ref{3.29}).
The commutation relations are also decomposed into
(\ref{3.37}), (\ref{3.38}), (\ref{3.39}), (\ref{3.43}) and (\ref{3.49}).
Noticeable effects of the central extension are
the anomalous commutator (\ref{3.43})
and the twist relation (\ref{3.49}).
These features also affect representation of the algebra
as seen in the following sections.
\subsection{Representations}
\subsubsection{Without the central extension}
Now we proceed to construct representations of the algebra
defined by (\ref{3.37})-(\ref{3.39}) and other vanishing commutators
with the constraints (\ref{3.29}), (\ref{3.30}) and (\ref{3.32}).
\par
Remember that $ \hat{P} $ and $ \hat{N} $ are self-adjoint
and that $ \hat{U} $ and $ \hat{W} $ are unitary.
Both of the relations (\ref{3.37}) and (\ref{3.38})
are isomorphic to (\ref{2.5}).
Hence the Ohnuki-Kitakado representations provide
representations for them, too.
$ \hat{P} $ and $ \hat{U} $ act on the Hilbert space $ H_\alpha $
via (\ref{2.9}) and (\ref{2.11}).
$ \hat{N} $ and $ \hat{W} $ act on another Hilbert space $ H_\beta $
via
\begin{eqnarray}
	&&
	\hat{N} \, | n + \beta \ket = ( n + \beta ) \, | n + \beta \ket,
	\label{3.51}
	\\
	&&
	\hat{W} \, | n + \beta \ket = | n + 1 + \beta \ket.
	\label{3.52}
\end{eqnarray}
The value of $ \alpha $ is arbitrary.
However $ \beta $ is restricted to be an integer
if we impose the condition (\ref{3.29}).
\par
 For $ \hat{\varphi} $ and $ \hat{\pi} $ the Fock representation works.
We define operators $ \hat{a}_n $ and $ \hat{a}_n^\dagger $ by
\begin{eqnarray}
	&&
	\hat{\varphi} (\sigma)
	=
	\frac{1}{2 \pi} \, \sum_{ n \ne 0 } \,
	\sqrt{ \frac{ \pi }{ | n | } } \,
	( \hat{a}_n         \, e^{   i n \sigma }
	+ \hat{a}_n^\dagger \, e^{ - i n \sigma } ),
	\label{3.53}
	\\
	&&
	\hat{\pi} (\sigma)
	=
	\frac{i}{2 \pi} \, \sum_{ n \ne 0 } \,
	\sqrt{ \pi | n | } \,
	( - \hat{a}_n         \, e^{   i n \sigma }
	  + \hat{a}_n^\dagger \, e^{ - i n \sigma } ).
	\label{3.54}
\end{eqnarray}
In the Fourier series the zero-mode $ n = 0 $ is excluded
because of the constraints (\ref{3.30}) and (\ref{3.32}).
It is easily verified that the commutator (\ref{3.39}) is equivalent to
\begin{equation}
	[ \, \hat{a}_m , \hat{a}_n^\dagger ] = \delta_{ m \, n }
	\;\;\;
	( m , n = \pm 1 , \pm 2 , \cdots )
	\label{3.55}
\end{equation}
with the other vanishing commutators.
Hence the ordinary Fock space $ F $ gives a representation
of $ \hat{a}^\dagger $'s and $ \hat{a} $'s,
which are called creation operators and annihilation operators, respectively.
\par
Consequently
the tensor product space $ H_\alpha \otimes H_0 \otimes F $ gives
an irreducible representation
of the fundamental algebra without the central extension.
The inequivalent ones are parametrized by $ \alpha $ $ ( 0 \le \alpha < 1 ) $.
\par
A remark is in order here;
the coefficients in front of $ \hat{a} $'s in (\ref{3.53}) and (\ref{3.54})
are chosen to diagonalize the hamiltonian of free field
\begin{eqnarray}
	\hat{H}
	& = &
	\frac12
	\int_0^{ 2 \pi }
	\left[
	\Bigl( \frac{1}{ 2 \pi } \hat{P} + \hat{\pi}(\sigma) \Bigr)^2
	+
	\Bigl( \partial \hat{\varphi}(\sigma) + \hat{N} \Bigr)^2
	\right]
	d \sigma
	\nonumber
	\\
	& = &
	\frac12
	\Bigl( \frac{1}{ 2 \pi } \hat{P}^2 + 2 \pi \hat{N}^2 \Bigr)
	+
	\sum_{ n \ne 0 } | n |
	\Bigl( \hat{a}_n^\dagger \, \hat{a}_n + \frac12 \Bigr).
	\label{3.56}
\end{eqnarray}
This hamiltonian corresponds to the lagrangian density
\begin{equation}
	{\cal L} =
	\frac{1}{2} \, \partial_\mu \phi^\dagger \, \partial^{\, \mu} \phi,
	\label{3.57}
\end{equation}
where $ \phi $ is given by (\ref{3.15}).
This lagrangian describes a massless boson.
Although Coleman's theorem~\cite{Coleman} forbids existence of massless bosons
in (1+1) dimensions due to infrared catastrophe,
our model is still permitted.
In our model the base space $ S^1 $ is compact,
hence there is no infrared divergence.
Interacting field theory will be briefly discussed later.
\par
Another remark is added.
The Kamefuchi-O'Raifeartaigh-Salam theorem~\cite{Kamefuchi} states that
the $ S $-matrix in quantum field theories remains unchanged
under any point transformation of field variables.
Their theorem is proved
within the framework of the conventional canonical formalism.
Thus this theorem is not applicable to our model.
If we take a real scalar field $ \varphi $ such that
$ \phi(\sigma) = e^{ i \varphi (\sigma) } $,
the lagrangian (\ref{3.57}) becomes a genuine free scalar theory,
\begin{equation}
	{\cal L} =
	\frac{1}{2} \, \partial_\mu \varphi \, \partial^{\, \mu} \varphi.
	\label{3.57a}
\end{equation}
In this case the Fock space is a unique representation,
thus there is no room for such an undetermined parameter $ \alpha $.
They considered only field theories with a trivial topology,
then they derived the equivalence theorem for $ S $-matrix.
On the other hand, we considered a field theory with a nontrivial topology,
then we reached existence of inequivalent representations.
However the $ S $-matrix of our model is not yet calculated,
hence it is left undetermined whether $ S $-matrix does depend on
the parameter $ \alpha $ or not.
\subsubsection{With the central extension}
Next we shall construct representations of the algebra
defined by
(\ref{3.37}), (\ref{3.38}), (\ref{3.39}), (\ref{3.43}), (\ref{3.49})
and other vanishing commutators
with the constraints (\ref{3.29}), (\ref{3.30}) and (\ref{3.32}).
The way of construction is a bit modified from to the previous one.
\par
Taking account of the twist relation (\ref{3.49}),
the representation of
$ \hat{P} $, $ \hat{U} $, $ \hat{N} $ and $ \hat{W} $ are given by
\begin{eqnarray}
	&&
	\hat{P}          \, | \, p + \alpha ; \, n \ket
	= ( p + \alpha ) \, | \, p + \alpha ; \, n \ket,
	\label{3.58}
	\\
	&&
	\hat{U} \, | \, p     + \alpha ; \, n \ket
	=          | \, p + 1 + \alpha ; \, n \ket,
	\label{3.59}
	\\
	&&
	\hat{N} \, | \, p + \alpha ; \, n \ket
	=    n     | \, p + \alpha ; \, n \ket,
	\label{3.60}
	\\
	&&
	\hat{W} \, | \, p       + \alpha ; \, n     \ket
	=          | \, p - 2 k + \alpha ; \, n + 1 \ket.
	\label{3.61}
\end{eqnarray}
The inner product is defined by
\begin{equation}
	\bra p + \alpha ; \, m \, | \, q + \alpha ; \, n \ket
	=
	\delta_{ p \, q } \, \delta_{ m \, n }
	\;\;\;
	( p , q , m , n \in \Z ).
	\label{3.62}
\end{equation}
The Hilbert space formed by completing the space of linear combinations of
$ | \, p + \alpha ; \, n \ket $
is denoted by $ T_\alpha $.
($ T $ indicates ``twist''.)
\par
Let us turn to $ \hat{\varphi} $ and $ \hat{\pi} $.
Considering the anomalous commutator (\ref{3.43}),
after a tedious calculation we obtain a Fourier expansion
\begin{eqnarray}
	&&
	\hat{\varphi} (\sigma)
	=
	\sum_{ n \ne 0 } \,
	\frac{1}{ \sqrt{ 2 |kn| } } \,
	( \hat{a}_n         \, e^{   i n \sigma }
	+ \hat{a}_n^\dagger \, e^{ - i n \sigma } ),
	\label{3.63}
	\\
	&&
	\hat{\pi} (\sigma)
	=
	\frac{i}{2 \pi}
	\left\{
	       \begin{array}{ll}
	       \sum_{ n = 1 }^{ \infty } & (k>0) \\
	       \sum_{ n =-1 }^{-\infty } & (k<0)
	       \end{array}
	\right\}
	\sqrt{ 2 |k n| }
	( - \hat{a}_n         \, e^{   i n \sigma }
	  + \hat{a}_n^\dagger \, e^{ - i n \sigma } ),
	\label{3.64}
\end{eqnarray}
where $ \hat{a} $'s and $ \hat{a}^\dagger $'s
obey the same commutation relations (\ref{3.55}).
The derivation of the above expansion is shown in the appendix.
It should be noticed that
only positive $n$'s appear in the expansion of $ \hat{\pi} $
when $ k > 0 $,
while only negative $n$'s appear
when $ k < 0 $.
However both positive and negative $ n $'s appear in $ \hat{\varphi} $.
Physical implication of lack of half of modes in $ \hat{\pi}(\sigma) $
is still unclear but will be discussed later.
\par
The algebra defined by (\ref{3.55})
is also represented by the Fock space $ F $.
Hence the tensor product space $ T_\alpha \otimes F $ gives
an irreducible representation
of the fundamental algebra with the central extension
for each value of $ \alpha \, ( 0 \le \alpha < 1 ) $.
\subsection{Normal ordering}
Although most of our main subjects are finished,
a subtle problem is still left.
At (\ref{3.31}) exponential of the local operator $ \hat{\varphi}(\sigma) $
is introduced.
As mentioned there, it contains divergence thus it is ill-defined.
Now we shall consider this problem closely.
\subsubsection{Without the central extension}
$ e^{ i \hat{\varphi}(\sigma) } $ should be expressed in terms of
creation and annihilation operators to act on the Fock space.
First we study the case without central extension.
If we define
\begin{equation}
	\hat{\chi} (\sigma)
	=
	\frac{1}{2 \pi} \, \sum_{ n \ne 0 } \,
	\sqrt{ \frac{ \pi }{ | n | } } \,
	\hat{a}_n  \, e^{ i n \sigma },
	\label{3.65}
\end{equation}
(\ref{3.53}) is decomposed into
$ \hat{\varphi}(\sigma) = \hat{\chi}(\sigma) + \hat{\chi}^\dagger(\sigma) $.
A bit calculation shows that
\begin{eqnarray}
	[ \, \hat{\chi}(\sigma) , \hat{\chi}^\dagger(\sigma') \, ]
	& = &
	\frac{1}{4 \pi}
	\biggl\{
		\sum_{n=1}^\infty \frac{1}{n} \, e^{ in ( \sigma - \sigma' ) }
		+
		\sum_{n=1}^\infty \frac{1}{n} \, e^{-in ( \sigma - \sigma' ) }
	\biggr\}
	\nonumber \\
	& = &
	\frac{1}{4 \pi}
	\{
		\log ( 1 - e^{  i ( \sigma - \sigma' ) } )
		+
		\log ( 1 - e^{ -i ( \sigma - \sigma' ) } )
	\},
	\label{3.66}
\end{eqnarray}
which diverges when $ \sigma' $ closes to $ \sigma $,
\begin{equation}
	\lim_{ \sigma' \to \sigma } \,
	[ \, \hat{\chi}(\sigma) , \hat{\chi}^\dagger(\sigma') \, ]
	=
	- \infty.
	\label{3.67}
\end{equation}
Thus application of the formula (\ref{3.44}) tells that
\begin{equation}
	e^{ i \hat{\varphi}(\sigma) }
	=
	e^{ i \hat{\chi}(\sigma) + i \hat{\chi}^\dagger(\sigma) }
	=
	e^{-\frac12 [ \, \hat{\chi}(\sigma) , \hat{\chi}^\dagger(\sigma) \, ] }
	\, e^{ i \hat{\chi}^\dagger(\sigma) }
	\, e^{ i \hat{\chi}(\sigma) }
	\label{3.68}
\end{equation}
diverges.
To eliminate the divergence we introduce the normal ordering procedure,
which is a rule to rearrange creation operators to the left
and annihilation operators to the right for each term.
This procedure is denoted by sandwiching by double colons, for example
\begin{equation}
	: e^{ i \hat{\varphi}(\sigma) } :  \,
	=
	e^{ i \hat{\chi}^\dagger(\sigma) } \,
	e^{ i \hat{\chi}(\sigma) }.
	\label{3.69}
\end{equation}
As (\ref{3.44}) implies that
$ e^{\hat{X}} \, e^{\hat{Y}} =
  e^{[ \, \hat{X}, \hat{Y} \, ]} \, e^{\hat{Y}} \, e^{\hat{X}} $,
(\ref{3.66}) gives
\begin{eqnarray}
	&&
	: e^{ i \hat{\varphi}(\sigma ) } :
	: e^{ i \hat{\varphi}(\sigma') } :
	\nonumber \\
	& = &
	e^{ i \hat{\chi}^\dagger(\sigma) }  \,
	e^{ i \hat{\chi}(\sigma) }          \,
	e^{ i \hat{\chi}^\dagger(\sigma') } \,
	e^{ i \hat{\chi}(\sigma') }
	\nonumber \\
	& = &
	e^{ -[ \, \hat{\chi}(\sigma) , \hat{\chi}^\dagger(\sigma') \, ] } \,
	e^{ i \hat{\chi}^\dagger(\sigma) }  \,
	e^{ i \hat{\chi}^\dagger(\sigma') } \,
	e^{ i \hat{\chi}(\sigma) }          \,
	e^{ i \hat{\chi}(\sigma') }
	\nonumber \\
	& = &
	\exp
	\left[
		- \frac{1}{4 \pi}
		\log ( 1 - e^{  i ( \sigma - \sigma' ) } )
		     ( 1 - e^{ -i ( \sigma - \sigma' ) } )
	\right]
	: e^{ i \hat{\varphi}(\sigma ) } \,
	  e^{ i \hat{\varphi}(\sigma') } :
	\label{3.70}
\end{eqnarray}
which is well-defined except for $ \sigma = \sigma' $
and invariant under permutation of $ \sigma $ with $ \sigma' $.
Therefore we conclude that
\begin{equation}
	: e^{ i \hat{\varphi}(\sigma ) } :
	: e^{ i \hat{\varphi}(\sigma') } :
	\, = \,
	: e^{ i \hat{\varphi}(\sigma') } :
	: e^{ i \hat{\varphi}(\sigma ) } :,
	\label{3.71}
\end{equation}
then (\ref{3.21}) is satisfied.
\par
The other relation (\ref{3.22}) also must be satisfied
after the normal ordering procedure.
Let us check it.
We substitute (\ref{3.54}) and
\begin{equation}
	f(\sigma) = \sum_{n \ne 0} f_n e^{ in\sigma }
	\;\;\;
	( f_{-n} = f_n^* )
	\label{3.72}
\end{equation}
into (\ref{3.33}) to obtain
\begin{equation}
	\hat{V} ( f )
	=
	\exp
	\left[
		- i \int_0^{2 \pi} f(\sigma) \hat{\pi}(\sigma) d \sigma
	\right]
	=
	\exp
	\left[
		\sum_{n \ne 0} \sqrt{ \pi |n| } \,
		( -f_{-n} \, \hat{a}_n + f_n \, \hat{a}_n^\dagger )
	\right].
	\label{3.73}
\end{equation}
Then it is easily seen that
\begin{eqnarray}
	&&
	\hat{V}^\dagger(f) \, \hat{a}_n \, \hat{V}(f)
	=
	\hat{a}_n + \sqrt{ \pi |n| } \, f_{n},
	\label{3.74}
	\\
	&&
	\hat{V}^\dagger(f) \, \hat{a}_n^\dagger \, \hat{V}(f)
	=
	\hat{a}_n^\dagger + \sqrt{ \pi |n| } \, f_{-n}.
	\label{3.75}
\end{eqnarray}
The definitions (\ref{3.65}) and (\ref{3.72}) with the above equations yield
\begin{eqnarray}
	\hat{V}^\dagger(f) \, \hat{\chi}(\sigma) \, \hat{V}(f)
	& = &
	\frac{1}{2 \pi} \, \sum_{ n \ne 0 } \,
	\sqrt{ \frac{ \pi }{ | n | } } \,
	( \hat{a}_n + \sqrt{ \pi |n| } \, f_{n} )
	 e^{   i n \sigma }
	\nonumber \\
	& = &
	\hat{\chi}(\sigma) + \frac12 f(\sigma),
	\label{3.76}
	\\
	\hat{V}^\dagger(f) \, \hat{\chi}^\dagger(\sigma) \, \hat{V}(f)
	& = &
	\frac{1}{2 \pi} \, \sum_{ n \ne 0 } \,
	\sqrt{ \frac{ \pi }{ | n | } } \,
	( \hat{a}_n^\dagger + \sqrt{ \pi |n| } \, f_{-n} )
	 e^{ - i n \sigma }
	\nonumber \\
	& = &
	\hat{\chi}(\sigma) + \frac12 f(\sigma),
	\label{3.77}
\end{eqnarray}
therefore
\begin{eqnarray}
	\hat{V}^\dagger(f) : e^{i\hat{\varphi}(\sigma)} : \hat{V}(f)
	& = &
	\hat{V}^\dagger(f)              \,
	e^{i\hat{\chi}^\dagger(\sigma)} \,
	e^{i\hat{\chi}(\sigma)}         \,
	\hat{V}(f)
	\nonumber \\
	& = &
	e^{if(\sigma)}                  \,
	e^{i\hat{\chi}^\dagger(\sigma)} \,
	e^{i\hat{\chi}(\sigma)}
	\nonumber \\
	& = &
	e^{if(\sigma)}
	: e^{i\hat{\varphi}(\sigma)} :.
	\label{3.78}
\end{eqnarray}
This result coincides with (\ref{3.22}).
Consequently we have checked that
the fundamental relations (\ref{3.21}) and (\ref{3.22}) are preserved
by the normal ordering procedure.
Although $ \hat{\phi}(\sigma) $ is claimed to be a unitary operator
above (\ref{3.21}),
$ \hat{\phi}(\sigma) = \, : e^{ i \hat{\varphi}(\sigma) } : $
is {\it not} unitary.
If it were unitary,
$ \hat{\phi}(\sigma) \, \hat{\phi}^\dagger(\sigma) $
must be equal to identity.
But actually
\begin{eqnarray}
	\lim_{\sigma' \to \sigma}
	\hat{\phi}(\sigma) \, \hat{\phi}^\dagger(\sigma')
	& = &
	\lim_{\sigma' \to \sigma}
	e^{ i\hat{\chi}^\dagger(\sigma) } \,
	e^{ i\hat{\chi}(\sigma)}          \,
	e^{-i\hat{\chi}^\dagger(\sigma')} \,
	e^{-i\hat{\chi}(\sigma')}         \,
	\nonumber \\
	& = &
	\lim_{\sigma' \to \sigma}
	e^{ [ \, \hat{\chi}(\sigma) , \hat{\chi}^\dagger(\sigma') \, ] } \,
	e^{ i\hat{\chi}^\dagger(\sigma) } \,
	e^{-i\hat{\chi}^\dagger(\sigma')} \,
	e^{ i\hat{\chi}(\sigma)}          \,
	e^{-i\hat{\chi}(\sigma')}         \,
	\nonumber \\
	& = &
	\lim_{\sigma' \to \sigma}
	e^{ [ \, \hat{\chi}(\sigma) , \hat{\chi}^\dagger(\sigma) \, ] } \,
	\hat{1}
	\nonumber \\
	& = &
	0
	\label{3.79}
\end{eqnarray}
due to (\ref{3.67}), hence
$ \hat{\phi}(\sigma) = \, : e^{ i \hat{\varphi}(\sigma) } : $
is not unitary.
\subsubsection{With the central extension}
Similarly we can verify the case of $ k \ne 0 $.
The way of verification is almost identical to the previous one
but a bit changed.
We define
\begin{equation}
	\hat{\chi} (\sigma)
	=
	\sum_{ n \ne 0 }
	\frac{1}{ \sqrt{2|kn|} } \,
	\hat{a}_n  \, e^{ i n \sigma },
	\label{3.80}
\end{equation}
to decompose (\ref{3.63}) into
$ \hat{\varphi}(\sigma) = \hat{\chi}(\sigma) + \hat{\chi}^\dagger(\sigma) $.
A bit calculation yields
\begin{eqnarray}
	[ \, \hat{\chi}(\sigma) , \hat{\chi}^\dagger(\sigma') \, ]
	& = &
	\frac{1}{2|k|}
	\biggl\{
		\sum_{n=1}^\infty \frac{1}{n} \, e^{ in ( \sigma - \sigma' ) }
		+
		\sum_{n=1}^\infty \frac{1}{n} \, e^{-in ( \sigma - \sigma' ) }
	\biggr\}
	\nonumber \\
	& = &
	\frac{1}{2|k|}
	\{
		\log ( 1 - e^{  i ( \sigma - \sigma' ) } )
		+
		\log ( 1 - e^{ -i ( \sigma - \sigma' ) } )
	\},
	\label{3.81}
\end{eqnarray}
which is also divergent when $ \sigma = \sigma' $.
To eliminate the divergence we use the normal ordering procedure again;
\begin{equation}
	: e^{ i \hat{\varphi}(\sigma) } : \,
	=
	e^{ i \hat{\chi}^\dagger(\sigma) } \,
	e^{ i \hat{\chi}(\sigma) }.
	\label{3.82}
\end{equation}
A calculation similar to (\ref{3.70}) yields
\begin{equation}
	: e^{ i \hat{\varphi}(\sigma ) } :
	: e^{ i \hat{\varphi}(\sigma') } : \,
	=
	\exp
	\left[
		- \frac{1}{2|k|}
		\log ( 1 - e^{  i ( \sigma - \sigma' ) } )
		     ( 1 - e^{ -i ( \sigma - \sigma' ) } )
	\right]
	: e^{ i \hat{\varphi}(\sigma ) }
	  e^{ i \hat{\varphi}(\sigma') } :.
	\label{3.83}
\end{equation}
Thus the commutativity (\ref{3.21}) is ensured again.
\par
Verification of the other relation (\ref{3.22}) is a bit complicated.
We define
\begin{eqnarray}
	&&
	f^{(+)}(\sigma) = \sum_{n= 1}^{ \infty} f_n \, e^{ in\sigma },
	\nonumber \\
	&&
	f^{(-)}(\sigma) = \sum_{n=-1}^{-\infty} f_n \, e^{ in\sigma },
	\nonumber \\
	&&
	f(\sigma) = f^{(+)}(\sigma) + f^{(-)}(\sigma).
	\;\;\;
	( f_{-n} = f_n^* )
	\label{3.84}
\end{eqnarray}
Substitution of (\ref{3.64}) into (\ref{3.50}) gives
\begin{eqnarray}
	\hat{V} ( f )
	& = &
	\exp
	\left[
		- i \int_0^{2 \pi} f(\sigma) \, \hat{\pi}(\sigma) \, d \sigma
	\right]
	\nonumber \\
	& = &
	\exp
	\left[
		\left\{
			\begin{array}{ll}
			\sum_{n= 1}^{ \infty} & (k>0) \\
			\sum_{n=-1}^{-\infty} & (k<0)
			\end{array}
		\right\}
		\sqrt{2|kn|}
		( -f_{-n} \, \hat{a}_n + f_n \, \hat{a}_n^\dagger )
	\right].
	\label{3.85}
\end{eqnarray}
It follows that
\begin{eqnarray}
	&&
	\hat{V}^\dagger(f) \, \hat{a}_n \, \hat{V}(f)
	=
	\hat{a}_n + \theta(kn) \sqrt{2|kn|} \, f_{n},
	\label{3.86}
	\\
	&&
	\hat{V}^\dagger(f) \, \hat{a}_n^\dagger \, \hat{V}(f)
	=
	\hat{a}_n^\dagger + \theta(kn) \sqrt{2|kn|} \, f_{-n},
	\label{3.87}
\end{eqnarray}
where $ \theta(x) $ is 1 when $ x > 0 $ and 0 when $ x < 0 $.
The definitions (\ref{3.80}) and (\ref{3.84}) with the above equations yield
\begin{eqnarray}
	\hat{V}^\dagger(f) \, \hat{\chi}(\sigma) \, \hat{V}(f)
	& = &
	\sum_{ n \ne 0 }
	\frac{1}{ \sqrt{2|kn|} }
	( \hat{a}_n + \theta(kn) \sqrt{2|kn|} \, f_{n} )
	e^{ i n  \sigma }
	\nonumber \\
	& = &
	\hat{\chi}(\sigma) + f^{(\pm)}(\sigma),
	\label{3.88}
	\\
	\hat{V}^\dagger(f) \, \hat{\chi}^\dagger(\sigma) \, \hat{V}(f)
	& = &
	\sum_{ n \ne 0 }
	\frac{1}{ \sqrt{2|kn|} }
	( \hat{a}_n^\dagger + \theta(kn) \sqrt{2|kn|} \, f_{-n} )
	e^{-i n  \sigma }
	\nonumber \\
	& = &
	\hat{\chi}^\dagger(\sigma) + f^{(\mp)}(\sigma),
	\label{3.89}
\end{eqnarray}
where the alternative sign is chosen according to the sign of $ k $.
Again we arrive at the same result
\begin{eqnarray}
	\hat{V}^\dagger(f) : e^{i\hat{\varphi}(\sigma)} : \hat{V}(f)
	& = &
	\hat{V}^\dagger(f)              \,
	e^{i\hat{\chi}^\dagger(\sigma)} \,
	e^{i\hat{\chi}(\sigma)}         \,
	\hat{V}(f)
	\nonumber \\
	& = &
	e^{if^{(+)}(\sigma) + if^{(-)}(\sigma)} \,
	e^{i\hat{\chi}^\dagger(\sigma)}         \,
	e^{i\hat{\chi}(\sigma)}
	\nonumber \\
	& = &
	e^{if(\sigma)}
	: e^{i\hat{\varphi}(\sigma)} :.
	\label{3.90}
\end{eqnarray}
\newpage
%
%
\section{Summary and discussion}
\subsection{Summary}
Now let us summarize what have been done in this paper.
We have reviewed the quantum mechanics on $ S^1 $ originally formulated
by Ohnuki and Kitakado.
They chose generators respecting topology of $ S^1 $.
They defined the algebra and classified its irreducible representations.
Inequivalent representations are characterized by a continuous parameter
$ \alpha $ $ ( 0 < \alpha \le 1 ) $.
Elimination of $ \alpha $ is obstructed by nontrivial topology of $ S^1 $.
\par
As a generalization of the quantum mechanics on $ S^1 $,
we have proposed the definition of the algebra for the abelian sigma model
in (1+1) dimensions.
The central extensions are also introduced into the algebra.
\par
The degrees of freedom of the field variable are separated as
\begin{equation}
	\map( S^1; S^1 )
	\cong
	S^1 \times \Z \times \map_0 ( S^1; \R ).
	\label{4.1}
\end{equation}
The separation is done as follows.
Identify $ \phi : S^1 \to S^1 $ with $ \phi : S^1 \to U(1) $.
Take a branch of its logarithm and put
$ \tilde{\varphi}(\sigma) = -i \log \phi(\sigma) $.
Define $ N \in \Z $
by $ 2 \pi N = \tilde{\varphi}(2\pi) - \tilde{\varphi}(0) $.
Next define $ e^{i \lambda} \in S^1 $ by
$ 2 \pi \lambda
= \int_0^{2\pi} ( \tilde{\varphi}(\sigma) - N \sigma ) d \sigma $.
 Finally define $ \varphi : S^1 \to \R $
by $ \varphi(\sigma) = \tilde{\varphi}(\sigma) - \lambda - N \sigma $.
Thus
$ \varphi(2\pi) = \varphi(0) $ and
$ \int_0^{2\pi} \varphi(\sigma) d \sigma = 0 $.
Then
\begin{equation}
	\phi(\sigma)
	=
	e^{ i ( \lambda + N\sigma + \varphi(\sigma) ) }
	\label{4.2}
\end{equation}
is a decomposition according to (\ref{4.1}) and results in (\ref{3.15})
by putting $ U = e^{ i \lambda } $.
What we have done is to define a coordinate system
in the infinite dimensional manifold $ Q = \map(S^1; S^1) $.
In the same way we can define a coordinate in the group
$ \Gamma = \map(S^1; U(1)) $ as given by (\ref{3.17}).
These coordinates are convenient;
they are direct product decompositions of the manifold $ Q $,
the group $ \Gamma $ and the action of $ \Gamma $ on $ Q $.
In other words,
these decompositions are preserved under group operation of $ \Gamma $
and are also preserved under the action of $ \Gamma $ on $ Q $
as shown in (\ref{3.18})-(\ref{3.20}).
\par
Existence of such coordinates is crucial
for construction of the quantum theory.
The fundamental relations (\ref{3.21})-(\ref{3.23})
are easy to understand intuitively,
however too complicated to construct its concrete representation.
The coordinates reduce them to simpler relations (\ref{3.37})-(\ref{3.39}).
Even if the central extension exists,
other complication is only addition of the anomalous commutator (\ref{3.43})
and the twist relation (\ref{3.49}).
Thus we have noticed that
the Ohnuki-Kitakado representations and the Fock representation provide
representations for our model.
\par
We conclude that inequivalent irreducible representations are parametrized
again by $ \alpha $ $ (0 \le \alpha <1) $.
When there is the central extension,
the action of the operator $ \hat{W} $ is changed as (\ref{3.61})
and
half of modes in $ \hat{\pi}(\sigma) $ is removed as (\ref{3.64}).
\par
Exponential of the local operator $ \hat{\varphi}(\sigma) $
must be regularized by the normal ordering procedure.
We have shown that the procedure preserves the fundamental relations
but violates unitarity of $ e^{ i \hat{\varphi}(\sigma) } $.
\subsection{Discussion}
 For what kind of physics is our theory applicable?
What we have done is just formulation of a rather ideal model.
It gives a lesson;
{\it when a model has nontrivial topology,
it is possible to construct inequivalent quantum theories,
even if they are equivalent as classical theories.}
We would like to point out some models which have such possibilities.
\par
The first example is still an ideal model in (1+1) dimensions;
it is the sine-Gordon model, whose lagrangian is
\begin{equation}
	{\cal L}
	=
	\frac12
	\partial_\mu      \tilde{\varphi}(x)
	\partial^{\, \mu} \tilde{\varphi}(x)
	+
	\kappa^2    \cos( \tilde{\varphi}(x) ).
	\label{4.3}
\end{equation}
It is a model which has interaction.
It can be rewritten by the variables of the abelian sigma model
without the central extension
by identifying $ \phi $ with $ e^{ i \tilde{\varphi} } $.
Then the corresponding hamiltonian is defined by
\begin{eqnarray}
	\hat{H}
	& = &
	\frac12
	\int_0^{ 2 \pi }
	\left[
	\Bigl( \frac{1}{ 2 \pi } \hat{P} + \hat{\pi} \Bigr)^2
	+
	\partial \hat{\phi}^\dagger \, \partial \hat{\phi}
	-
	\kappa^2 ( \hat{\phi} + \hat{\phi}^\dagger )
	\right]
	d \sigma
	\nonumber
	\\
	& = &
	\frac12
	\Bigl( \frac{1}{ 2 \pi } \hat{P}^2 + 2 \pi \hat{N}^2 \Bigr)
	+
	\sum_{ n \ne 0 } | n |
	\Bigl( \hat{a}_n^\dagger \, \hat{a}_n + \frac12 \Bigr)
	\nonumber
	\\
	&&
	-
	\frac{\kappa^2}{2}
	\int_0^{ 2 \pi } \Bigl(
		\hat{U}
		e^{  i ( \hat{N}\sigma + \hat{\varphi}(\sigma) ) }
		+
		\hat{U}^\dagger
		e^{ -i ( \hat{N}\sigma + \hat{\varphi}(\sigma) ) }
	\Bigr) d \sigma.
	\label{4.4}
\end{eqnarray}
The last term contains highly nonlinear complicated interaction.
It also contains interaction between the zero-mode $ \hat{U} $
and the fluctuation mode $ \hat{\varphi} $.
This hamiltonian commutes with $ \hat{N} $,
hence the winding number is conserved.
If the hamiltonian includes the winding operator $ \hat{W} $,
change of the winding number can occur.
Such a jumping motion is not allowed in classical theory
but is possible in quantum theory.
The winding number is sometimes called soliton or kink number.
It is known~\cite{soliton} that
this model has a topological soliton, which behaves like a fermion.
However it is still obscure
whether our formulation is relevant to soliton physics or not.
It is expected that our model
may give an insight to quantum theory of solitons.
\par
Other examples are found in both field theories and string theories.
 From both points of view, it is hoped to extend our model
to nonabelian groups and to higher dimensions.
Our model has the field configuration space $ Q = \map( S^1; S^1 ) $.
The group $ \Gamma = \map( S^1; U(1) ) $ acts on $ Q $ simply transitively.
The most general model has $ Q = \map( X; M ) $,
in which $ X $ is called a base space and $ M $ is a target space.
When $ \dim X > 1 $, we call it a higher dimensional model.
If a group $ G $ acts on $ M $ transitively,
$ M $ is called a homogeneous space $ G/H $.
Then an infinite dimensional group $ \Gamma = \map( X; G ) $ acts transitively
on $ Q = \map( X; M ) $ by pointwise multiplication.
When the group $ G $ is nonabelian, it is called a nonlinear sigma model.
When the base space $ X $ is $ S^1 $,
it is called a bosonic string model.
In addition, when the target space $ M $ is an orbifold,
for example a toroidal orbifold $ T^n/Z $,
it is called an orbifold string model.
Directions for extensions are summarized in the table.
\begin{table}
\caption{Possible extensions}
\vspace{4mm}
\begin{tabular}[h]{|l|l|l|l|}
\hline
        & abelian sigma model & higher dimensions  & nonabelian           \\
\hline
base space   $ X $ & $ S^1 $  & $ S^n, T^n, \R^n $ &                      \\

target space $ M $ & $ S^1 $  &                    & $ G, G/H, T^n/Z $    \\

group        $ G $ & $ U(1) $ &              & $ SU(n), SO(n) $ {\it etc} \\
\hline
\end{tabular}
\end{table}
\par
We should refer to the known results on nonlinear sigma models
in two dimensions.
Some of them are exactly solved by the method of factorization theory and
the Bethe ansatz~\cite{Exact}.
Here ``exactly solved'' means that the exact $ S $-matrix is obtained
and therefore the mass spectrum defined by poles of the $ S $-matrix
is also calculated.
Wiegmann {\it et al}~\cite{Wiegmann} have already obtained exact solutions
of nonlinear sigma models for the algebras
$ SO(n+2), \, SU(n+1), \, Sp(2n) \, ( n = 1, 2, \cdots ) $.
All of them exhibit massive spectra.
\par
Their approach is quite different from ours.
In fact they use neither field variables nor lagrangians.
They demand some reasonable properties to be satisfied by the $ S $-matrix;
unitarity, factorizability, crossing symmetry, analyticity
and other symmetries specified by a Lie algebra.
In (1+1) dimensions such a requirement determines the $ S $-matrix
directly and unambiguously.
Actually what they have constructed is a realization of symmetries
in terms of $ S $-matrix.
But they do not pay attention to topology.
Although the model which we have considered is quite simple,
that is the $ SO(2) $ sigma model,
we have shown existence of inequivalent quantizations
as a consequence of the nontrivial topology.
At present we do not know how to incorporate topology
into the algebraic approach of Zamolodchikov and Wiegmann {\it et al}.
\par
We shall briefly comment upon the known results
on orbifold string models~\cite{Narain}.
Sakamoto {\it et al}~\cite{Sakamoto} investigated
theories of closed bosonic strings on orbifolds in operator formalism.
He have shown
that commutators among the zero-mode and winding-mode variables
are left ambiguous
and that these variables obey nontrivial quantization.
The relation between his result and ours is still left obscure.
\par
 Finally we would like to suggest a way to explore
nonabelian and higher dimensional theories.
The decomposition (\ref{4.1}) heavily relies
on the abelian nature of the group $ U(1) $.
As a generalization to a nonabelian group $ G $, we expect a decomposition
\begin{equation}
	\Gamma
	= \map( S^n ; G )
	\cong
	G \times \pi_n(G) \times \map_0 ( S^n ; \mbox{Lie}(G) ),
	\label{4.5}
\end{equation}
where $ \pi_n $ denotes the $ n $-th homotopy group and
\begin{equation}
	\map_0 ( S^n ; \mbox{Lie}(G) )
	=
	\{
		\, g : S^n \to \mbox{Lie}(G)
		\, | \,
		C^\infty, \,
		\int_{S^n} g = 0
	\}.
	\label{4.6}
\end{equation}
In the decomposition (\ref{4.5}),
the first component describes the zero-mode,
the second one does the topologically disconnected mode,
and the third one does the fluctuation mode.
Unfortunately such a decomposition does not exist,
because the nonabelian nature severely entangles
degrees of freedom of $ \Gamma $.
It seems hopeless to find a convenient coordinate in $ \Gamma $ generally.
We should take a rather abstract approach to construct a representation
for such a complicated group.
Gelfand {\it et al}~\cite{Gelfand} has investigated
the representation theory of
the group $ \Gamma = \map( X; G ) $ for
a base space $ X (\dim X \ge 2) $ and a compact semisimple $ G $.
They do not rely on any coordinates but take a quite abstract approach.
However they do not consider a manifold $ Q = \map( X; G/H ) $
on which $ \Gamma $ acts.
We do not know how to incorporate such a configuration space $ Q $
into their representation theory.
%
%
\section*{Acknowledgments}
I am indebted to so many people
that I cannot record all of their names here.
I gratefully acknowledge helpful discussions with
T.Kashiwa and S.Sakoda (Kyushu University),
K.Higashijima (Osaka University),
M.Sakamoto and M.Tachibana (Kobe University),
I.Tsutsui and S.Tsujimaru (Institute for Nuclear Study),
K.Fujii (Yokohama City University)
and Y.Igarashi (Niigata University).
I wish to express my gratitude to Y.Ohnuki, S.Kitakado and H.Ikemori,
who discussed with me and encouraged me continuously.
I thank all members of E-laboratory,
who provided me with encouraging environment
during the whole period of my studying.
%
%
\newpage
\appendix
\section{Appendix}
Here we give an explicit calculation of the Fourier expansion of
(\ref{3.39}) and (\ref{3.43})
to derive (\ref{3.63}), (\ref{3.64}) and (\ref{3.55}).
We repeat the assumptions;
\begin{eqnarray}
	&&
	[ \, \hat{\varphi} ( \sigma ) , \hat{\varphi} ( \sigma' ) \, ]
	=
	0,
	\label{A.1}
	\\
	&&
	[ \, \hat{\varphi} ( \sigma ) , \hat{\pi} ( \sigma' ) \, ]
	=
	i \Bigl( \delta( \sigma - \sigma' ) - \frac{1}{2 \pi} \Bigr),
	\label{A.2}
	\\
	&&
	[ \, \hat{\pi} (\sigma) , \hat{\pi} (\sigma') \, ]
	=
	- \, \frac{i k}{\pi} \, \delta' ( \sigma - \sigma' ),
	\label{A.3}
\end{eqnarray}
with constraints (\ref{3.30}) and (\ref{3.32}).
$ k $ is assumed to be a non-zero integer.
We define
\begin{eqnarray}
	&&
	\hat{\varphi}_m
	=
	\frac{1}{2\pi}
	\int_0^{2\pi} \hat{\varphi}(\sigma) e^{- i m \sigma} d \sigma,
	\label{A.4}
	\\
	&&
	\hat{\pi}_n
	=
	\frac{1}{2\pi}
	\int_0^{2\pi} \hat{\pi}(\sigma) e^{ - i n \sigma } d \sigma,
	\label{A.5}
\end{eqnarray}
for integers $ m, n $.
Obviously,
$ \hat{\varphi}^\dagger_m = \hat{\varphi}_{-m} $,
$ \hat{\pi}^\dagger_n     = \hat{\pi}_{-n}     $ and
$ \hat{\varphi}_0         = \hat{\pi}_0    = 0 $.
Multiplying $ (2\pi)^{-2} \, e^{-i m \sigma -i n \sigma'} $
to (\ref{A.1})-(\ref{A.3})
and integrating
$ \int_0^{2\pi} \int_0^{2\pi} d \sigma \, d \sigma' $,
we obtain
\begin{eqnarray}
	&&
	[ \, \hat{\varphi}_m , \hat{\varphi}_n \, ]
	=
	0,
	\label{A.6}
	\\
	&&
	[ \, \hat{\varphi}_m , \hat{\pi}_n \, ]
	=
	\frac{i}{2\pi}
	( \delta_{m+n, 0} - \delta_{m,0} \delta_{n,0} ),
	\label{A.7}
	\\
	&&
	[ \, \hat{\pi}_m , \hat{\pi}_n \, ]
	=
	\frac{k}{2 \pi^2} \, m \, \delta_{m+n, 0}.
	\label{A.8}
\end{eqnarray}
We put
\begin{equation}
	\hat{b}_n =
	 i \sqrt{ \frac{ 2\pi^2 }{ |k| n } } \, \hat{\pi}_n,
	\qquad\qquad
	\hat{b}_n^\dagger =
	-i \sqrt{ \frac{ 2\pi^2 }{ |k| n } } \, \hat{\pi}_{-n},
	\;\;\;
	( n > 0 )
	\label{A.9}
\end{equation}
then (\ref{A.8}) implies
\begin{eqnarray}
	&&
	[ \, \hat{b}_m , \hat{b}_n \, ] = 0,
	\label{A.10}
	\\
	&&
	[ \, \hat{b}_m , \hat{b}_n^\dagger \, ]
	= \epsilon (k) \, \delta_{m n},
	\qquad
	( m, n > 0 )
	\label{A.11}
\end{eqnarray}
where $ \epsilon(k) = k / |k| $.
By definition we have
\begin{equation}
	\hat{\pi}_n
	=
	-i \sqrt{ \frac{ |k|n }{ 2\pi^2 } } \, \hat{b}_n,
	\qquad\qquad
	\hat{\pi}_{-n}
	=
	 i \sqrt{ \frac{ |k|n }{ 2\pi^2 } } \, \hat{b}_n^\dagger,
	\qquad
	( n > 0 )
	\label{A.12}
\end{equation}
hence
\begin{eqnarray}
	\hat{\pi} (\sigma)
	& = &
	\sum_{ n \ne 0 } \hat{\pi}_n \, e^{ i n \sigma}
	\nonumber \\
	& = &
	\frac{i}{2 \pi} \sum_{ n = 1 }^{ \infty }
	\sqrt{ 2 |k| n } \,
	( - \hat{b}_n         \, e^{   i n \sigma }
	  + \hat{b}_n^\dagger \, e^{ - i n \sigma } ).
	\label{A.13}
\end{eqnarray}
On the other hand, for non-zero $ m $ and $ n $,
(\ref{A.6}) and (\ref{A.7}) yield
\begin{equation}
	[ \,
	\hat{\varphi}_m - \frac{i \pi}{km} \hat{\pi}_m ,
	\hat{\pi}_n
	\, ]
	=
	\frac{i}{2 \pi} \delta_{m+n,0}
	- \frac{i \pi}{km} \, \frac{km}{2 \pi^2} \, \delta_{m+n,0}
	= 0
	\label{A.14}
\end{equation}
and also
\begin{equation}
	[ \,
	\hat{\varphi}_m - \frac{i \pi}{km} \hat{\pi}_m ,
	\hat{\varphi}_n - \frac{i \pi}{kn} \hat{\pi}_n
	\, ]
	=
	\frac{1}{2kn} \delta_{m+n,0}.
	\label{A.15}
\end{equation}
Therefore if we put
\begin{equation}
	\left\{
	\begin{array}{l}
		\hat{b}_{-n}
		=
		\epsilon(k) \sqrt{ 2 |k| n } \,
		( \hat{\varphi}_{-n} + \frac{i \pi}{kn} \hat{\pi}_{-n} ),
		\\
		\hat{b}_{-n}^\dagger
		=
		\epsilon(k) \sqrt{ 2 |k| n } \,
		( \hat{\varphi}_{ n} - \frac{i \pi}{kn} \hat{\pi}_{ n} ),
	\end{array}
	\right.
	\;\;\;
	( n > 0 )
	\label{A.16}
\end{equation}
they satisfy
\begin{eqnarray}
	&&
	[ \, \hat{b}_m , \hat{b}_n \, ] = 0,
	\label{A.17}
	\\
	&&
	[ \, \hat{b}_m , \hat{b}_n^\dagger \, ]
	= \epsilon (k) \, \delta_{m n}.
	\qquad
	( m, n \ne 0 )
	\label{A.18}
\end{eqnarray}
It is easily seen that
\begin{eqnarray}
	&&
	\hat{\varphi}_n
	=
	\frac{ \epsilon(k) }{ \sqrt{2|k|n} } \hat{b}_{-n}^\dagger
	+ \frac{i \pi}{kn} \hat{\pi}_n
	=
	\frac{ \epsilon(k) }{ \sqrt{2|k|n} }
	( \hat{b}_n + \hat{b}_{-n}^\dagger ),
	\label{A.19}
	\\
	&&
	\hat{\varphi}_{-n}
	=
	\frac{ \epsilon(k) }{ \sqrt{2|k|n} } \hat{b}_{-n}
	- \frac{i \pi}{kn} \hat{\pi}_{-n}
	=
	\frac{ \epsilon(k) }{ \sqrt{2|k|n} }
	( \hat{b}_{-n} + \hat{b}_n^\dagger ),
	\;\;\;
	(n>0)
	\label{A.20}
\end{eqnarray}
using (\ref{A.16}) with (\ref{A.12}).
Thus we have
\begin{eqnarray}
	\hat{\varphi} (\sigma)
	& = &
	\sum_{ n \ne 0 } \hat{\varphi}_n \, e^{ i n \sigma}
	\nonumber \\
	& = &
	\epsilon(k)
	\sum_{ n \ne 0 }
	\frac{1}{ \sqrt{ 2 |k n| } }
	( \hat{b}_n         \, e^{   i n \sigma }
	+ \hat{b}_n^\dagger \, e^{ - i n \sigma } ).
	\label{A.21}
\end{eqnarray}
 Finally we define
\begin{equation}
	\hat{a}_n = \left\{
			    \begin{array}{l}
			      \hat{b}_n            \\
			    - \hat{b}_{-n}^\dagger
			    \end{array}
		     \right.
	\;\;\;
	\hat{a}_n^\dagger = \left\{
				    \begin{array}{ll}
				      \hat{b}_n^\dagger & (k>0) \\
				    - \hat{b}_{-n}      & (k<0)
			            \end{array}
		             \right.
	\label{A.22}
\end{equation}
to get
\begin{eqnarray}
	&&
	[ \, \hat{a}_m , \hat{a}_n^\dagger \, ] = \delta_{mn},
	\label{A.23}
	\\
	&&
	\hat{\varphi} (\sigma)
	=
	\sum_{ n \ne 0 }
	\frac{1}{ \sqrt{ 2 |k n| } }
	( \hat{a}_n         \, e^{   i n \sigma }
	+ \hat{a}_n^\dagger \, e^{ - i n \sigma } ),
	\label{A.24}
	\\
	&&
	\hat{\pi} (\sigma)
	=
	\frac{i}{2 \pi}
	\left\{
	       \begin{array}{ll}
	       \sum_{ n = 1 }^{ \infty } & (k>0) \\
	       \sum_{ n =-1 }^{-\infty } & (k<0)
	       \end{array}
	\right\}
	\sqrt{ 2 |k n| }
	( - \hat{a}_n         \, e^{   i n \sigma }
	  + \hat{a}_n^\dagger \, e^{ - i n \sigma } ),
	\label{A.25}
\end{eqnarray}
which are the desired results (\ref{3.55}), (\ref{3.63}) and (\ref{3.64}).
\newpage

%
\end{document}